\documentclass[reprint,onecolumn,notitlepage,preprintnumbers,nofootinbib,amsmath,amssymb,aps,prd,floatfix,superscriptaddress,longbibliography,svgnames]{revtex4-1}

\usepackage{amsmath}
\usepackage{amsfonts}
\usepackage{amssymb}
\usepackage{graphicx}
\usepackage{bm}
\usepackage{subfigure}
\usepackage{url}
\usepackage[hyperindex]{hyperref}
\usepackage{color}
\usepackage[ddmmyy,24hr]{datetime}
\usepackage{bigdelim}
\usepackage{booktabs}
\usepackage{dcolumn}
\usepackage{multirow}
\usepackage{subfigure}
\usepackage{physics}
\usepackage{cancel}
\usepackage{stackrel}
\usepackage{paralist}
\usepackage{xspace}
\usepackage{slashed}
\usepackage{cancel}
\usepackage{todonotes}
\usepackage{enumerate}
\usepackage{float}
\usepackage{fullpage}
\usepackage{ulem}
\usepackage{wasysym}
\usepackage{comment}
\usepackage{bbold}
\usepackage{hyperref}

\usepackage{xcolor}
\usepackage{orcidlink}

\usepackage{feynmp-auto}
\newcommand{\nua}[1]{\ensuremath{\rlap{\kern-2.5pt\ensuremath{\overset{\scriptscriptstyle(-)}{\phantom{\nu}}}}{\ensuremath{{\nu}_{#1}}}}}

\usepackage{enumitem}

\newcommand{\cenns}{CE$\nu$NS }
\newcommand{\nues}{$\nu $ES }

\begin{document}
\title{Momentum dependent flavor radiative corrections to the coherent elastic neutrino-nucleus scattering for the neutrino charge-radius determination}

\author{M. Atzori Corona \orcidlink{0000-0001-5092-3602}}
\email{mattia.atzori.corona@ca.infn.it}
\affiliation{Dipartimento di Fisica, Universit\`{a} degli Studi di Cagliari,
	Complesso Universitario di Monserrato - S.P. per Sestu Km 0.700,
	09042 Monserrato (Cagliari), Italy}
\affiliation{Istituto Nazionale di Fisica Nucleare (INFN), Sezione di Cagliari,
	Complesso Universitario di Monserrato - S.P. per Sestu Km 0.700,
	09042 Monserrato (Cagliari), Italy}

\author{M. Cadeddu \orcidlink{0000-0002-3974-1995}}
\email{matteo.cadeddu@ca.infn.it}
\affiliation{Istituto Nazionale di Fisica Nucleare (INFN), Sezione di Cagliari,
	Complesso Universitario di Monserrato - S.P. per Sestu Km 0.700,
	09042 Monserrato (Cagliari), Italy}

\author{N. Cargioli \orcidlink{0000-0002-6515-5850}}
\email{nicola.cargioli@ca.infn.it}
\affiliation{Dipartimento di Fisica, Universit\`{a} degli Studi di Cagliari,
	Complesso Universitario di Monserrato - S.P. per Sestu Km 0.700,
	09042 Monserrato (Cagliari), Italy}
\affiliation{Istituto Nazionale di Fisica Nucleare (INFN), Sezione di Cagliari,
	Complesso Universitario di Monserrato - S.P. per Sestu Km 0.700,
	09042 Monserrato (Cagliari), Italy}

\author{F. Dordei \orcidlink{0000-0002-2571-5067}}
\email{francesca.dordei@cern.ch}
\affiliation{Istituto Nazionale di Fisica Nucleare (INFN), Sezione di Cagliari,
	Complesso Universitario di Monserrato - S.P. per Sestu Km 0.700,
	09042 Monserrato (Cagliari), Italy}

\author{C. Giunti \orcidlink{0000-0003-2281-4788}}
\email{carlo.giunti@to.infn.it}
\affiliation{Istituto Nazionale di Fisica Nucleare (INFN), Sezione di Torino, Via P. Giuria 1, I--10125 Torino, Italy}

\date{\dayofweekname{\day}{\month}{\year} \ddmmyydate\today, \currenttime}

\begin{abstract}
Despite being neutral particles, neutrinos can have a non-zero charge radius, which represents the only non-null neutrino electromagnetic property in the standard model theory. Its value can be predicted with high accuracy and its effect is usually accounted for through the definition of a radiative correction affecting the neutrino couplings to electrons and nucleons at low energy, which results effectively in a shift of the weak mixing angle. Interestingly, it introduces a flavour-dependence in the cross-section. Exploiting available neutrino-electron and coherent elastic neutrino-nucleus scattering (CE$\nu$NS) data, there have been many attempts to measure experimentally the neutrino charge radius. Unfortunately, the current precision allows one to only determine constraints on its value.
In this work, we discuss how to properly account for the neutrino charge radius in the \cenns cross-section including the effects of the non-null momentum-transfer in the neutrino electromagnetic form factor, which have been usually neglected when deriving the aforementioned limits. We apply the formalism discussed to a re-analysis of the COHERENT cesium iodide and argon samples and the NCC-1701 germanium data from the Dresden-II nuclear power plant. We quantify the impact of this correction on the CE$\nu$NS cross-section and we show that, despite being small, it can not be neglected in the analysis of data from future high-precision experiments. Furthermore, this momentum dependence can be exploited to significantly reduce the allowed values for the neutrino charge radius determination. 
\end{abstract}

\maketitle  

\section{Introduction}
Over the years, the Standard Model (SM) electroweak theory has been carefully studied, both theoretically and experimentally, to accurately determine the couplings between interacting particles. The quest for increasingly higher precision brought to the introduction of the so-called radiative corrections~\cite{Erler:2013xha}, due to higher-order vertex contributions.
Available formalisms have been developed such that most of these radiative corrections are rather universal in neutral current interactions at low energy, meaning that they do not depend on the particular particles involved in the process. In this work, we discuss neutrino interactions, focusing in particular on the coherent elastic neutrino-nucleus scattering (CE$\nu$NS) process~\cite{Freedman:1973yd}. In this case, another non-universal radiative correction has to be accounted for, the so-called neutrino charge radius (NCR) radiative correction. The latter represents the only flavor-dependent contribution to the CE$\nu$NS cross-section, which otherwise would be a completely flavor-blind process.\\

Despite the experimental CE$\nu$NS era having just started~\cite{COHERENT:2017ipa}, the precision already achieved in the current measurements~\cite{COHERENT:2021xmm,COHERENT:2020iec, Colaresi:2022obx} permits to put stringent constraints on different beyond the SM (BSM) quantities (see e.g. Refs.~\cite{AtzoriCorona:2022qrf,AtzoriCorona:2023ktl,Cadeddu:2023tkp}), requiring thus a careful treatment of the radiative corrections affecting this specific process.
In this work, we revisit the canonical description of the neutrino charge radius radiative correction to account for its momentum dependence, due to the non-null momentum transfer in the neutrino electromagnetic form factor, which has been so far neglected in CE$\nu$NS experimental and phenomenological studies. We quantify the impact of this correction and we later apply it to the available \cenns data analyses, namely the COHERENT cesium iodide (CsI)~\cite{COHERENT:2021xmm} and argon (Ar)~\cite{COHERENT:2020iec} samples and the NCC-1701 germanium (Ge) data from the Dresden-II nuclear power plant~\cite{Colaresi:2022obx}. Nevertheless, the formalism discussed in this work would impact also other low-energy neutrino interactions, like e.g. neutrino elastic scattering processes on electrons ($\nu$ES). 

\section{Theoretical framework}
\label{sec:cs}

\subsection{Radiative corrections to the coherent elastic neutrino-nucleus scattering cross section}
Low-energy neutrino elastic scattering processes on both nuclei and electrons are of fundamental interest in modern electroweak physics. The SM theory well describes the cross-section for both processes.
Focusing on the SM CE$\nu$NS interactions, the differential cross section as a function of the nuclear kinetic recoil energy $T_\mathrm{nr}$ for a neutrino $\nu_{\ell}$ ($\ell=e,\mu,\tau$) scattering off a spin-zero nucleus $\mathcal{N}$ with $Z$ protons and $N$ neutrons is given by~\cite{Freedman:1973yd,Drukier:1984vhf,Barranco:2005yy,Patton:2012jr,Hoferichter:2020osn}
\begin{equation}
	\dfrac{d\sigma_{\nu_{\ell}\text{-}\mathcal{N}}}{d T_\mathrm{nr}}
	(E,T_\mathrm{nr})
	=
	\dfrac{G_{\text{F}}^2 M}{\pi}
	\left( 1 - \dfrac{M T_\mathrm{nr}}{2 E^2} \right)
	\left[g_{V}^{p}\left(\nu_{\ell}\right) Z F_{Z}\left(|\vec{q}|^{2}\right)+g_{V}^{n} N F_{N}\left(|\vec{q}|^{2}\right)\right]^{2},
	\label{cs-std}
\end{equation}
where $G_{\text{F}}$ is the Fermi constant, $E$ is the neutrino energy and $M$ the nuclear mass, while $F_{Z}\left(|\vec{q}|^{2}\right)$ and $F_{N}\left(|\vec{q}|^{2}\right)$ are the proton and neutron form factors, respectively, which account for the contribution of the nuclear structure to the process. In particular, they represent the Fourier transform of the nuclear proton and neutron density distributions, so that they account for the spatial distribution of nucleons inside the target nucleus. In Eq.~(\ref{cs-std}), $g_V^p$ and $g_V^n$ are the neutrino couplings to protons and neutrons, respectively, whose tree-level values are given by
\begin{align}
	g_{V}^{p} = \frac{1}{2}-2\;\textrm{sin}^2\;\vartheta_{\text{W}} \simeq 0.0227,\quad\quad
	g_{V}^{n} = -\frac{1}{2}\,,
	\label{gVpnSM}
\end{align}
where $\vartheta_{\text{W}}$ is the so-called weak mixing angle, also known as the Weinberg angle, a key parameter of the electroweak theory whose value evaluated at zero-momentum transfer is equal to $\textrm{sin}^2\vartheta_{\text{W}}(q^2=0)=\hat{s}^2_0=0.23863(5)$~\cite{ParticleDataGroup:2020ssz}. These values are evaluated neglecting all the higher-order vertex corrections to the CE$\nu$NS process. More precise values are determined by taking into account the radiative corrections in the Minimal Subtraction ($\overline{\mathrm{MS}}$) scheme, following Refs.~\cite{Erler:2013xha,ParticleDataGroup:2020ssz,AtzoriCorona:2023ktl}, with a well-defined
subtraction of singular terms arising in dimensional regularization, giving rise to expressions with a logarithmic energy scale dependence that is governed by a renormalization group equation (RGE). Choosing this energy scale equal to the
momentum transfer of the process under consideration will, in general, avoid spurious logarithms~\cite{PhysRevD.72.073003}. Thus, accounting for radiative corrections, the couplings become
\begin{align}
g_V^{p}(\nu_\ell)&=\rho\left(\frac{1}{2}-2\;\textrm{sin}^2\vartheta_{\text{W}}\right)+2\; {\raisebox{-1pt}{\rotatebox{90}{\Bowtie}}}_{WW}+\Box_{WW}-2\phi_{\nu_\ell W}+\rho(2\boxtimes_{ZZ}^{uL}+\boxtimes_{ZZ}^{dL}-2\boxtimes_{ZZ}^{uR}-\boxtimes_{ZZ}^{dR})\label{eq:gvp} \\
g_V^{n}&=-\frac{\rho}{2}+2\Box_{WW}+{\raisebox{-1pt}{\rotatebox{90}{\Bowtie}}}_{WW}+\rho(2\boxtimes_{ZZ}^{dL}+\boxtimes_{ZZ}^{uL}-2\boxtimes_{ZZ}^{dR}-\boxtimes_{ZZ}^{uR}).\label{eq:gvn}
\end{align}
The quantities $\Box_{WW}$, ${\raisebox{-1pt}{\rotatebox{90}{\Bowtie}}}_{WW}$ and $\boxtimes_{ZZ}^{fX}$, are the radiative corrections associated with the so-called $WW$ box diagram, the $WW$ crossed-box and the $ZZ$ box respectively, while $\rho=1.00063$ represents a low-energy correction for neutral-current processes. These contributions have the forms
\begin{equation}\label{WWbox}
\Box_{WW} = - {\hat\alpha_Z \over 2 \pi \hat s_Z^2}\left[ 1 - {\hat\alpha_s(M_W) \over 2 \pi} \right], 
\qquad\qquad
\hspace{3pt}{\raisebox{-1pt}{\rotatebox{90}{\Bowtie}}}_{WW} = 
{\hat\alpha_Z \over 8 \pi \hat s_Z^2} \left[ 1 + {\hat\alpha_s(M_W) \over \pi} \right], 
\end{equation}
\begin{equation}\label{eq:ZZbox}
\boxtimes_{ZZ}^{fX} = - {3 \hat\alpha_Z \over 8 \pi \hat s_Z^2 \hat c_Z^2} (g_{LX}^{\nu_\ell f})^2
\left[ 1 - {\hat\alpha_s(M_Z) \over\pi} \right],
\end{equation}
where $M_W$ is the $W$ boson mass, $\hat{\alpha}_Z$ is the electromagnetic coupling constant evaluated at the $Z$ boson mass, $\hat{\alpha}_s$ is the strong coupling constant, $f\in\{u,d\}$ and $X\in\{L,R\}$ is the parity. Note that in Eq.~(\ref{eq:ZZbox}) all the $(g_{LX})^{\nu_\ell f}$ couplings are evaluated at the lowest order of perturbation theory and the weak mixing angle is evaluated at the $Z$-pole, i.e. $\textrm{sin}^2\vartheta_{\text{W}}(M_Z)=\hat{s}^2_Z=0.23122(4)$~\cite{ParticleDataGroup:2020ssz}.
The remaining radiative term in Eq.~(\ref{eq:gvp}), namely $\phi_{\nu_\ell W} $, is referred to as the neutrino charge radius contribution. In the SM, it is defined as~\cite{Erler:2013xha}
\begin{equation}
\phi_{\nu_\ell W} = - {\alpha \over 6 \pi} \left( \ln {M_W^2 \over m_\ell^2} + {3 \over 2} \right),\label{phiNCR}
\end{equation}
where $\alpha$ is the low-energy limit of the electromagnetic coupling, and $m_\ell$ is the mass of the charged lepton with flavour $\ell$. Clearly, this radiative contribution depends on the neutrino flavor, thus introducing a flavor dependence in the neutrino-proton coupling alone.
Numerically, the values of the couplings correspond to 
\begin{align}
	g_{V}^{p}(\nu_{e}) \simeq 0.0381,\quad\quad
	g_{V}^{p}(\nu_{\mu}) \simeq 0.0299,\quad\quad
	g_{V}^{p}(\nu_{\tau}) \simeq 0.0255,\quad\quad
	g_{V}^{n} \simeq -0.5117\,.
	\label{gVn}
\end{align}
In the following, we will analyse in detail the different formalisms used to derive the NCR correction and we will focus in particular on its momentum dependence.
\subsection{Neutrino charge radius radiative correction}
Neutrinos are neutral particles, therefore, they can not couple directly with photons. However, even if the electric charge is null, the electric form factor, $\mathbb{f}_Q(q^2)$, carries non-trivial information about the neutrino electric properties, like the neutrino charge radius, millicharge and magnetic moment. In principle, a neutral particle can be characterized by a superposition of two different charge distributions of opposite signs described by an electric form factor which is nonzero only for momentum transfers different from zero, $q^2\neq 0$~\cite{Giunti:2014ixa}.
Expanding the form factor in a series of powers of $q^2$ we get
\begin{equation}
    \mathbb{f}_Q(q^2)=\mathbb{f}_Q(0)+q^2\dfrac{d\mathbb{f}_Q(q^2)}{dq^2}\Big|_{q^2=0}+\dots\, .\label{ElectricExpansion}
\end{equation}
In the ``Breit frame'', the charge form factor depends only on $|\vec{q}|$ and it can be interpreted as the Fourier transform of a spherically symmetric charge distribution, $\rho(r)$, so that~\cite{Giunti:2014ixa}
\begin{equation}
    \mathbb{f}_Q(q^2)=\int \rho(r) e^{-i\vec{q}\cdot \vec{r}}d^3r=\int \rho(r)\dfrac{\sin(qr)}{qr}d^3r\, .\label{electricRho}
\end{equation}
From this interpretation, it is easy to understand that the first term in the expansion in Eq.~(\ref{ElectricExpansion}), $\mathbb{f}_Q(0)$, has to be zero, since neutrinos are neutral, while the second term is identified as the neutrino charge radius, i.e. the radius of the electric charge distribution.
By deriving the expression in Eq.~(\ref{electricRho}) with respect to $q^2$ and taking the limit for $q^2\rightarrow 0$, we obtain
\begin{equation}
    \lim_{q^2\rightarrow 0}\dfrac{d\mathbb{f}_Q(q^2)}{dq^2}=\int \rho(r) \dfrac{r^2}{6}d^3r=\dfrac{\langle r^2\rangle}{6}\, ,
\end{equation}
where we introduced the squared neutrino charge radius $\langle r^2\rangle$ which is given by
\begin{equation}
    \langle r^2\rangle\equiv 6\dfrac{d\mathbb{f}_Q(q^2)}{dq^2}\Big|_{q^2=0}\, .\label{NCRdefDerivative}
\end{equation}
Let us note that $\langle r^2\rangle$ has no defined sign because $\rho(r)$ is not a positively defined quantity.\\
\begin{figure}[h]
\centering
\includegraphics[width=0.6\textwidth]{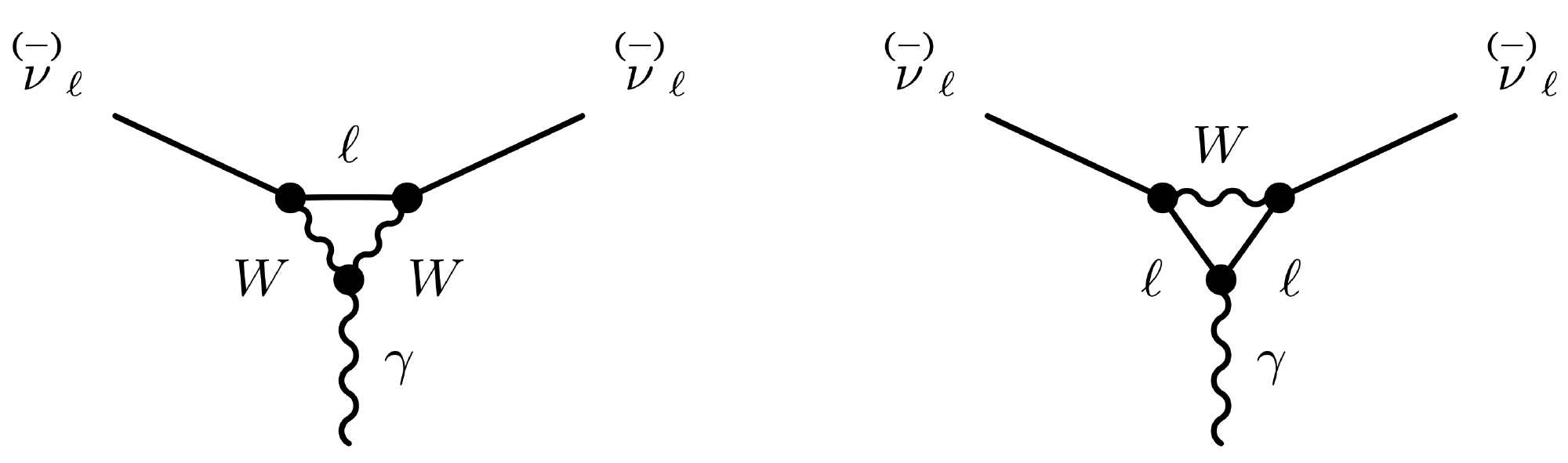}
\caption{Diagrams describing the neutrino charge radius contributions to the photon-neutrino vertices.}\label{DiagramNCR}
\end{figure}
Practically, the charge radius of a neutrino is generated by a loop insertion into the $\nu_\ell$ line, where $W$ boson(s) and charged lepton(s) $\ell$ can enter, as shown by the $WW\ell$ loop (left diagram) and the $\ell\ell W$ loop (right diagram) in Fig.~\ref{DiagramNCR}. These diagrams' contribution can be calculated, and according to Refs.~\cite{Bernabeu:2000hf,Bernabeu:2002nw,Bernabeu:2002pd} the NCR corresponds to a physical observable, being finite and gauge invariant. In particular, the SM calculation gives
\begin{equation}
    \langle r^2_{\nu_\ell}\rangle_{\rm SM}=-\dfrac{G_F}{2\sqrt{2}\pi^2}\Big[3-2\ln \Big(\dfrac{m_\ell^2}{M_W^2}\Big)\Big]\, ,   \label{NCRdef}
\end{equation}
where the sum of two contributions appears. Namely, one constant term arises from the $WW\ell$ loop diagram, where the photon interacts with the $W$ boson, and another is generated by the photon's interaction with the lepton $\ell$. The latter generates an electroweak logarithm, which is divergent in the UV range, that is regularized at the lepton mass, $m_\ell$, thus introducing a dependence of the NCR from the lepton flavor~\cite{Erler:2013xha,Bernabeu:2002pd}.
In the SM, the numerical values of the neutrino charge radius for the different neutrino flavors are
\begin{align}
    \langle r^2_{\nu_e}\rangle &\simeq-8.3\times 10^{-33}\, \mathrm{cm}^2\, ,\label{NCReeSM}\\
    \langle r^2_{\nu_\mu}\rangle &\simeq-4.8\times 10^{-33}\, \mathrm{cm}^2\, ,\label{NCRuuSM}\\
    \langle r^2_{\nu_\tau}\rangle &\simeq-3.0\times 10^{-33}\, \mathrm{cm}^2\, .\label{NCRttSM}
\end{align}
The neutrino charge radius affects the scattering of neutrinos with charged particles. Therefore, it contributes to the \nues process, whereas in the case of \cenns it contributes only to the neutrino-proton coupling, and not to the neutron one, as visible in Eqs.~(\ref{eq:gvp}) and~(\ref{eq:gvn}). There, the NCR contribution modifies the proton coupling constant~\cite{Giunti:2014ixa,AtzoriCorona:2022qrf,Cadeddu:2019eta}
\begin{equation}
    g_V^p(\nu_\ell)\rightarrow \tilde{g}_V^p-\dfrac{2}{3}M_W^2 \langle r^2_{\nu_\ell}\rangle\sin^2\vartheta_W= \tilde{g}_V^p-\dfrac{\sqrt{2}\pi\alpha}{3G_F}\langle r^2_{\nu_\ell}\rangle\, ,    \label{NCRradiative}
\end{equation}
where $\tilde{g}_V^p\simeq0.0184$ is the neutrino-proton coupling without the contribution of the SM NCR. The latter appears thus as a radiative correction, as shown in Eqs.~(\ref{eq:gvp}) and~(\ref{phiNCR}), that practically generates an effective shift of $\sin^2\!\vartheta_{W}$.
In the case of $\bar\nu_{\ell}\text{-}\mathcal{N}$ scattering, it is sufficient to operate the substitutions:
$g_{V}^{p,n} \to - g_{V}^{p,n}$ and $\langle{r}_{\nu_{\ell}}\rangle \to \langle{r}_{\bar\nu_{\ell}}\rangle = - \langle{r}_{\nu_{\ell}}\rangle$.
Therefore, the NCR of neutrinos and antineutrinos contribute with the same sign to the shift of the weak mixing angle in the CE$\nu$NS cross-section.\\

Despite a measurement of the NCR would represent a fundamental test of the SM, the available data are still insufficient to provide a first determination.
So far, only constraints have been put on its value~\cite{AtzoriCorona:2022qrf,DeRomeri:2022twg,TEXONO:2009knm,BNL,ParticleDataGroup:2020ssz}. One has to keep in mind that when considering BSM effects that affect the NCR, it is also possible to have so-called off-diagonal flavor-changing contributions, which are often referred to as transition charge radii. Using available data, limits on these quantities have been derived, see e.g. Ref.~\cite{AtzoriCorona:2022qrf}. Nevertheless, in the following, we will focus only on diagonal NCRs. In this way, we probe the values of the NCR in the SM. However, since it is also likely that BSM physics generates off-diagonal neutrino charge radii that are much smaller than the diagonal ones and that can thus be neglected in a first approximation, new physics models can also be tested in this scenario.\\

\begin{figure}
\centering
\subfigure[]{\label{fig:enter-label2}
\includegraphics[width=0.48\textwidth]{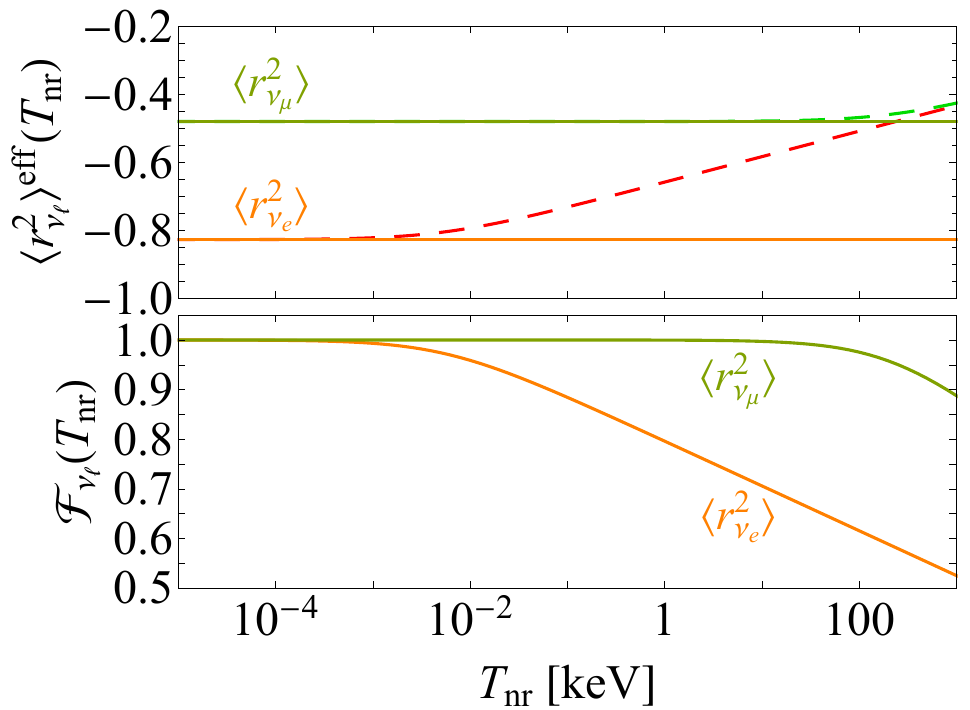}}
\subfigure[]{\label{fig:gvpCoup}
\includegraphics*[width=0.483\textwidth]{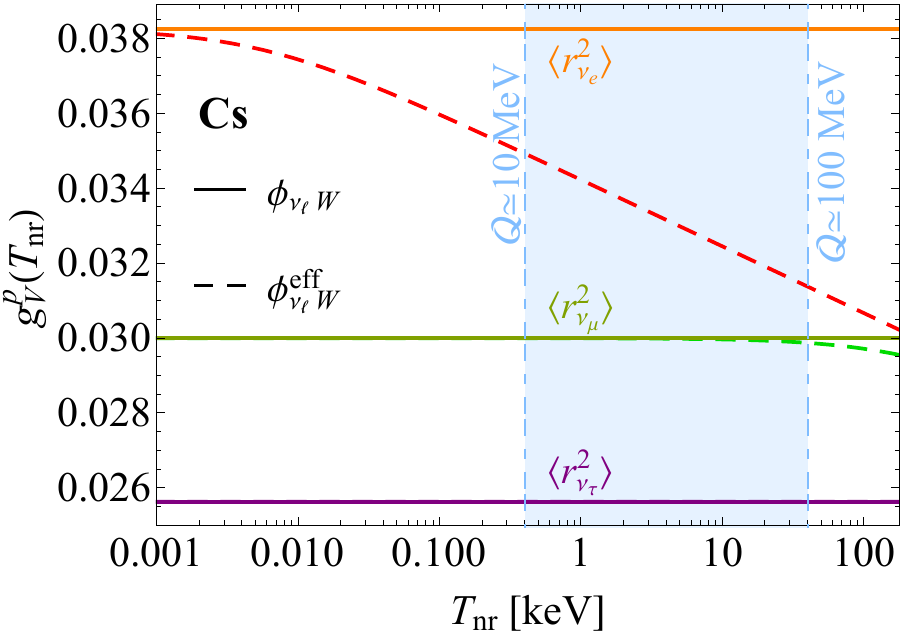}
}
\caption{ (a) (Upper) Comparison between the value of the NCR at zero-momentum transfer (solid lines) and the effective definition (dashed lines) in the case of the muonic flavor (green/lime) and the electron one (orange/red) as a function of the nuclear recoil energy for a cesium target. (Lower) Effective form factor as a function of the nuclear recoil energy for the muon flavor (green) and the electron one (orange).
(b) Neutrino-proton coupling for the CE$\nu$NS process on cesium nuclei as a function of the nuclear recoil energy. The solid lines refer to the couplings considering a constant NCR radiative correction, while the dashed ones to the momentum-dependent neutrino charge radius radiative correction case. The vertical light blue lines and the shaded area indicate the typical momentum transfers of CE$\nu$NS experiments.}
\end{figure}

Only recently, the RGE formalism was developed to make the weak mixing angle value not dependent on the particular scattering particle, such that the NCR is incorporated as an additional radiative correction to the coupling constant as shown in Eq.~(\ref{NCRradiative})~\cite{PhysRevD.72.073003,Erler:2017knj}.
Historically, the NCR was considered as one of the radiative corrections contributing to the running of the weak mixing angle in the case of neutrino scattering processes~\cite{Marciano:2003eq,SEHGAL1985370}. 
In this scenario, the running is evaluated only at the one-loop level using an effective form factor $k_{\nu_\ell}(q^2)$ which allows one to define the weak mixing angle at a certain energy scale starting from its value at the $Z$-mass energy scale, namely
\begin{equation}
    \sin^2\vartheta_W(q^2)= k_{\nu_\ell}(q^2)\sin^2\vartheta_W(M_Z)\equiv  k_{\nu_\ell}(q^2)\hat{s}^2_Z\, ,\label{wmaRunMarciano}
\end{equation}
where $ k_{\nu_\ell}(q^2)$ is defined by
\begin{align}\nonumber
    k_{\nu_\ell}(q^2)&=1-\dfrac{\alpha}{2\pi \hat{s}^2_Z}\Big[2\sum_f (T_{3f}Q_f-2\hat{s}^2_ZQ_f^2) J_f(q^2)+
    \dfrac{\hat{c}^2_Z}{3}+\dfrac{1}{\hat{c}^2_Z}\Big(\dfrac{19}{8}+\dfrac{17}{4}\hat{s}^2_Z+3\hat{s}_Z^4\Big)\\
    &-\Big(\dfrac{7}{2}\hat{c}_Z^2+\dfrac{1}{12}\Big)\ln \hat{c}^2_Z\Big]-\dfrac{\alpha}{\pi \hat{s}^2_Z}\Big[-R_\ell(q^2)+\dfrac{1}{4}\Big]\, ,\label{knuell}
\end{align}
and we defined $\hat{s}^2_Z\equiv \sin^2\vartheta_W(M_Z)$ and $\hat{c}^2_Z\equiv \cos^2\vartheta_W(M_Z)$. $T_{3f}$ is the fermion weak isospin third component, $Q_f$ is the fermion charge, and the sum runs over all the fermions $f$. It is crucial to notice that while the $J_f(q^2)$ term is relative to all the fermions $f$, the $R_\ell(q^2)$ one depends on the lepton flavor $\ell$ of the neutrino involved in the scattering process. These two terms are defined by the following integrals
\begin{align}
    J_f(q^2)&=\int_0^1 dx\, x(1-x)\ln\Big[\dfrac{m_f^2-q^2x(1-x)}{M_Z^2}\Big]\, ,\\
    R_\ell(q^2)&=\int_0^1 dx\, x(1-x)\ln\Big[\dfrac{m_\ell^2-q^2x(1-x)}{M_W^2}\Big]\, ,\label{RellDef}
\end{align}
showing a very similar structure but for the different masses in the denominator. 
If no neutrino is involved in the scattering process, the last term in Eq.~(\ref{knuell}), which involves also $R_\ell(q^2)$, vanishes. In this case one retrieves the definition of the weak mixing angle running for a generic scattering process, which can thus be compared with the RGE running (see also Ref.~\cite{PhysRevD.72.073003}) in the zero momentum limit, obtaining
\begin{equation}
    k_{\nu_\ell}(q^2=0)\hat{s}^2_Z-\hat{s}_0^2(\mathrm{RGE})=-\dfrac{2\alpha}{9\pi}+\mathcal{O}(\alpha^2)\, .
\end{equation}
It can be noticed that the difference of the weak mixing angle values consists only in a constant term (see also Appendix A of Ref.~\cite{Davoudiasl:2023cnc}).
Therefore, the last term in Eq.~(\ref{knuell}), in the $q^2\rightarrow 0$ limit, reduces to the NCR flavor-dependent radiative correction as defined in the RGE formalism (see Eq.~(\ref{phiNCR})), which can thus be also written as
\begin{equation}
    \phi_{\nu_\ell W}=-\dfrac{\alpha}{\pi}\left(-R_\ell(0)+\dfrac{1}{4}\right) \quad\quad \mathrm{with} \quad \quad R_\ell(0)=\dfrac{1}{6}\ln \dfrac{m_\ell^2}{M_W^2}\, .
\end{equation}
However, the interesting aspect of the latter definition is that it can be easily extended to consider the case of a nonzero momentum transfer just by letting the momentum in the $R_\ell$ integral be different from zero. This represents a more realistic scenario since all the experiments are run at a nonzero momentum transfer. By doing so, we are explicitly distinguishing between the NCR radiative correction and the actual physical NCR. In fact, the NCR is by definition independent of the momentum transfer, as it represents the value of the form factor derivative at zero momentum transfer, as shown in Eq.~(\ref{NCRdefDerivative}), while the NCR radiative correction has a momentum dependence, as it enters as a correction to the coupling. \\
Thus, we can introduce such a momentum dependence in the NCR radiative correction by defining
\begin{align}
    \phi_{\nu_\ell W}^\mathrm{eff}(q^2)&=-\dfrac{\alpha}{\pi}\left(-R_\ell(q^2)+\dfrac{1}{4}\right) \nonumber\\
    &=-\dfrac{\alpha}{\pi}\left(-\int_0^1dx\, x(1-x)\ln\Big[\dfrac{m_\ell^2-q^2x(1-x)}{M_W^2}\Big]+\dfrac{1}{4}\right)\, ,
\end{align}
which should then be used in the experimental extraction of the neutrino charge radius whenever \mbox{$q^2\neq0$}. Practically speaking, one should carefully correct for the momentum dependence, and this can be done introducing a ``neutrino charge radius form factor". We define this form factor by isolating the momentum-dependent NCR with respect to the SM picture, so basically as
\begin{equation}
    \mathcal{F}_{\nu_\ell}(T_{\rm nr})=\dfrac{\langle r^2_{\nu_\ell}\rangle^{\rm eff}(T_{\rm nr})}{\langle r^2_{\nu_\ell}\rangle^{\rm eff}(0)}\, \equiv \dfrac{\langle r^2_{\nu_\ell}\rangle^{\rm eff}(T_{\rm nr})}{\langle r^2_{\nu_\ell}\rangle^{\rm SM}}\, ,\label{FFncr}
\end{equation}
where we introduced an effective NCR\footnote{Since the NCR is a well-defined physical quantity, as also shown in Eq.~(\ref{NCRdefDerivative}), it is worth stressing that this effective definition has not to be considered as the actual definition of the neutrino charge radius. The charge radius is by definition evaluated at zero-momentum transfer. This effective definition can be thus viewed as a form factor which incorporates the momentum dependence, in analogy to what is commonly discussed for the nuclear form factor~\cite{AtzoriCorona:2023ktl}.}, namely
\begin{equation}
    \langle r_{\nu_\ell}^2\rangle^{\rm eff}=\dfrac{6 G_F}{\sqrt{2}\pi\alpha}\phi^{\rm eff}_{\nu_\ell W}(q^2)=-\dfrac{G_F}{2\sqrt{2}\pi^2}\Big[3-12R_\ell(q^2)\Big]\, ,
\end{equation}
so that, we can obtain the neutrino couplings by using this effective NCR radiative correction instead of the classical one inside Eq.~(\ref{NCRradiative}) and analogously for the case of the neutrino-electron coupling. 
In this manner, the NCR form factor is naturally normalized to unity for a zero-momentum transfer and decreases as the momentum grows, with a different steepness depending on the considered neutrino flavor. 
The momentum dependence of $\phi_{\nu_\ell W}^{\rm eff}(q^2)$ comes from $R_\ell(q^2)$ as defined in Eq.~(\ref{RellDef}), and its impact becomes relevant for momenta larger than the mass of the corresponding charged lepton $\ell$ that enters the loops in Fig.~\ref{DiagramNCR}, i.e. $q^2\gtrsim m_\ell^2$. Thus, for $\nu_e$ processes the correction to the couplings becomes visible for $q\gtrsim 0.5\, \mathrm{MeV}$, while for $\nu_\mu$ only above $\sim 100\, \mathrm{MeV}$. In the case of $\nu_\tau$ an even higher momentum transfer is needed to appreciate any difference, which is not relevant for the typical momenta of the $\nu$ES and CE$\nu$NS experiments that we are considering in this work. We remind here that the momentum transfer is derived as a function of the nuclear or electron recoil energy as
\begin{equation}
    Q^2=-q^2\simeq 2 m_{\rm tar} T\, ,
\end{equation}
where $m_{\rm tar}$ is the target mass, so either the nuclear or the electron mass, and $T$ is the nuclear, $T_{\rm nr}$, or electron, $T_{e}$, recoil energy, depending on the case.
In Fig.~\ref{fig:enter-label2}, we compare the SM value of the electron and muon NCRs with its effective value as a function of the nuclear recoil energy (considering a cesium nucleus). Moreover, in the lower panel of the figure we show the behaviour of the NCR form factor, $\mathcal{F}_{\nu_\ell}$, as a function of the nuclear recoil energy for both flavors.\\

We can now consider the effect of the NCR correction on the neutrino-proton coupling. In Fig.~\ref{fig:gvpCoup} we show the variation of the latter as a function of the nuclear recoil energy by comparing the two NCR radiative correction definitions. If we consider the typical momentum transfer (and thus recoil energy) of CE$\nu$NS experiments, $Q\sim10-100\, \mathrm{MeV}$, as shown by the shaded light blue area, we can notice that the variation of the coupling is clearly non-negligible for the $\nu_e$ case.
The effect on the $\nu_\mu$ proton coupling is, as expected, significant for higher energies, so that, for $\nu_\mu$ CE$\nu$NS at the current precision one can still employ a constant NCR radiative correction.
Quantitatively, considering the recoil energy region of interest for CE$\nu$NS, the variation of the $\nu_e$ proton coupling due to the NCR correction momentum dependence is between $\sim10-20\%$, while we see that there is almost no effect for the other two neutrino flavors.\\

The overall effect on the CE$\nu$NS cross section is small. Indeed, this effect involves mostly $\nu_e$, so that in the case of COHERENT only one flux component is affected.
Moreover, the variation affects only the coupling to protons, which is naturally suppressed by the weak mixing angle, so in practice, we are dealing with a minor effect. We estimate that the systematic bias of the $\nu_e-\mathcal{N}$ scattering cross section is around $1-2\%$, which is an effect of about 20\% with respect to the current systematic uncertainties affecting CE$\nu$NS for the CsI COHERENT dataset~\cite{COHERENT:2021xmm}.
For reference, the current statistical precision on the flux averaged CE$\nu$NS cross section measured by COHERENT is about 15-18\% for the CsI data~\cite{COHERENT:2021xmm} and about $\sim32\%$ for the Ar one~\cite{COHERENT:2020iec}. Nevertheless, for future measurements, it will become imperative to account for the momentum dependence of the NCR correction, especially as the community is putting a great effort into reaching the high precision frontier. Moreover, if one wants also to perform the first neutrino charge radius measurement, this effect will heavily mislead its extracted value, as the quantity that one measures, in particular in the case of the electron neutrinos, is the effective NCR, which has to be corrected for the momentum dependence in order to extract the physical NCR.\\

For completeness, we briefly discuss here also the comparison with the radiative corrections calculated in Ref.~\cite{Tomalak:2020zfh}, in which an Effective Field Theory (EFT) approach has been employed to determine them for the CE$\nu$NS process. The formalism is rather different with respect to the one considered in our work, however, we estimate that the two descriptions should be equivalent. In the EFT approach, the flavor-dependent contribution, given by the neutrino charge radius, has been evaluated through a polarization diagram, which is indeed equivalent to the integral defined in Eq.~(\ref{RellDef}). In Ref.~\cite{Tomalak:2020zfh}, the authors themselves compared the prescription commonly adopted in literature~\cite{Bernabeu:2002pd,Cadeddu:2018dux} with their EFT approach, considering however only the zero-momentum limit, so that a more complete and careful comparison is left for future studies.

\section{Results}
\label{sec:result}
In this section, we will present the results of the analysis of the diagonal neutrino charge radii, $\langle r^2_{\nu_{e}}\rangle$ and $\langle r^2_{\nu_{\mu}}\rangle$, obtained using the latest COHERENT cesium iodide and argon data-set, both alone and in combination, with the germanium NCC-1701 data from the Dresden-II nuclear reactor power plant\footnote{To model the antineutrino spectra from the Dresden-II reactor we have considered the so-called HMVE parametrization, obtained by combining the expected spectra for antineutrino energies above
2 MeV from Refs.~\cite{Mueller:2011nm, Huber:2011wv}, that we indicate as HM, with the low energy
part determined by Ref.~\cite{PhysRevD.39.3378}, that we indicate as VE. Nevertheless, the impact of the particular antineutrino flux on the results obtained is
negligible. Moreover, we use the Fef quenching factor~\cite{Colaresi:2022obx}.}. To analyse the data, we follow the same procedures described in Ref.~\cite{AtzoriCorona:2022qrf}. 
In particular, we focus on the comparison between the neutrino charge radii results obtained from CE$\nu$NS data that are present in literature (see e.g. Ref.~\cite{AtzoriCorona:2022qrf} for a strict comparison, given that the same formalism has been used, but also Ref.~\cite{DeRomeri:2022twg}), which are obtained considering the NCR correction without any momentum dependence (i.e. $\mathcal{F}_{\nu_\ell}\equiv1$), with our reanalysis in which we include the impact of a momentum-dependent form factor (i.e. $\mathcal{F}_{\nu_\ell}(T_{\rm nr})$). In the latter case, we are attempting to extract the physical value of the NCR at zero momentum transfer, whereas, neglecting the momentum dependence, one is measuring only an average effective neutrino charge radius relative to the typical momentum transfer of the experiment.
\begin{figure}[h]
    \centering
    \includegraphics[width=0.49\textwidth]{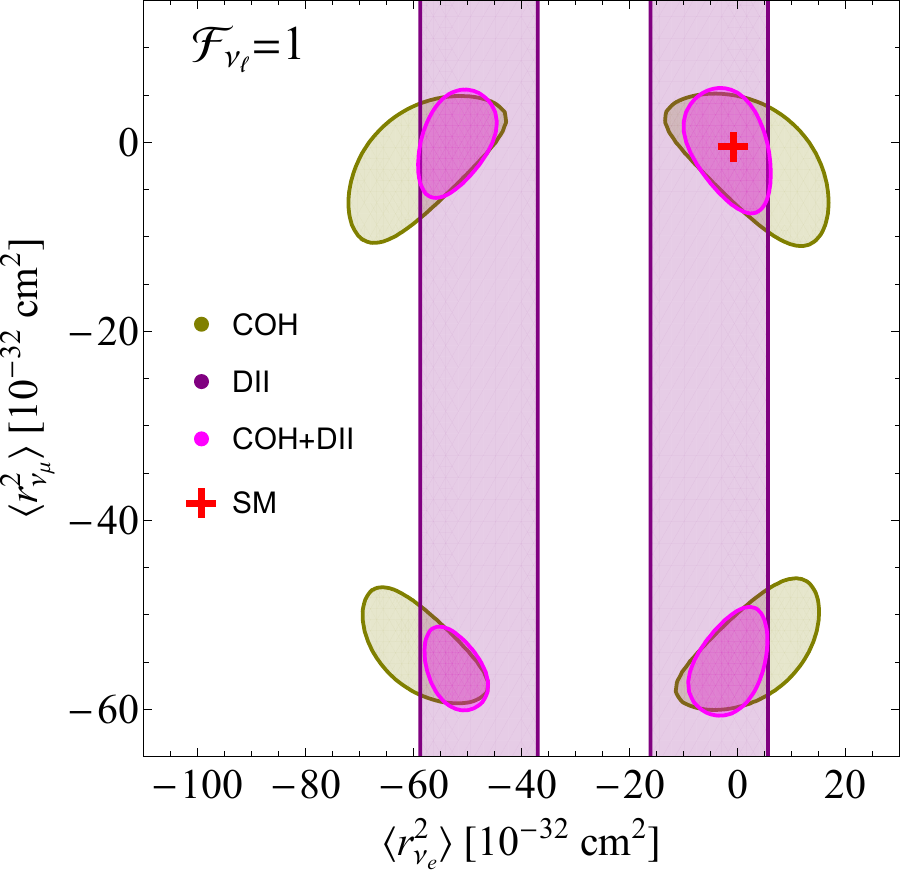}
    \includegraphics[width=0.49\textwidth]{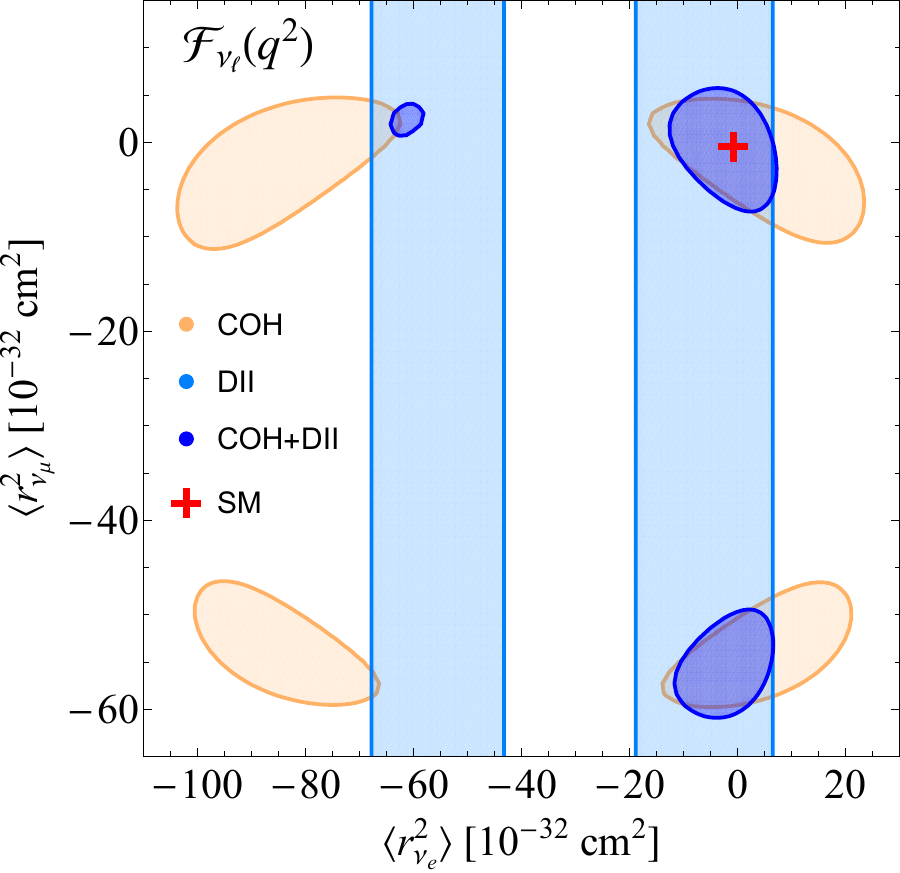}
    \caption{Allowed regions at 90\% CL from the analysis of the latest COHERENT CsI and Ar data (COH), Dresden-II (DII) and their combination (COH+DII) in the $\langle r^2_{\nu_{e}}\rangle$ vs $\langle r^2_{\nu_{\mu}}\rangle$ plane in case of momentum-independent (left) and -dependent (right) neutrino charge radii correction. The red cross indicates the SM values in Eqs.~(\ref{NCReeSM}) and~(\ref{NCRuuSM}).}
    \label{fig:COHCsIDII}
\end{figure}

By using the form factor $\mathcal{F}_{\nu_\ell}$ definition in Eq.~(\ref{FFncr}) and shown in Fig.~\ref{fig:enter-label2}, 
we assume that possible BSM contributions affect only the value of the neutrino charge radius defined at zero-momentum transfer and the momentum dependence of the NCR radiative correction is the same as in the SM.
Therefore, the neutrino-proton coupling defined in Eq.~(\ref{NCRradiative}) is modified by
\begin{equation}
    g_V^p(\nu_\ell,T_{\rm nr})\rightarrow \tilde{g}_V^p-\dfrac{\sqrt{2}\pi\alpha}{3G_F}\langle r^2_{\nu_\ell}\rangle\mathcal{F}_{\nu_\ell}(T_{\rm nr})\, ,
\end{equation}
where $\langle r^2_{\nu_\ell}\rangle$ is the physical value of the neutrino charge radius at zero-momentum transfer that we aim to measure through the data.\\

The results from the analysis of COHERENT CsI, Ar and Dresden-II Ge data are shown in Fig.~\ref{fig:COHCsIDII} at $90\%$ CL, where the left plot shows the $\mathcal{F}_{\nu_\ell}\equiv1$ limit, while in the right plot a momentum-dependent charge radius form factor $\mathcal{F}_{\nu_\ell}(T_{\rm nr})$ has been used. 
Since reactors provide only a flux of electron antineutrinos, the analysis of the Dresden-II data results in two degenerate bands, in correspondence of the SM $\langle r^2_{\nu_e}\rangle$ value and for a large negative one, which produces a degenerate value of the \cenns cross-section as defined in Eq.~(\ref{cs-std}). Instead, the COHERENT CsI and Ar data analyses produce 4 closed contours, since they are also sensitive to the muonic flavor. These allowed regions correspond to the SM values of the electron and muon neutrino charge radii and to large negative charge radius values which produce a degenerate cross-section.\\
\begin{figure}[h]
    \centering
\includegraphics[width=0.49\textwidth]{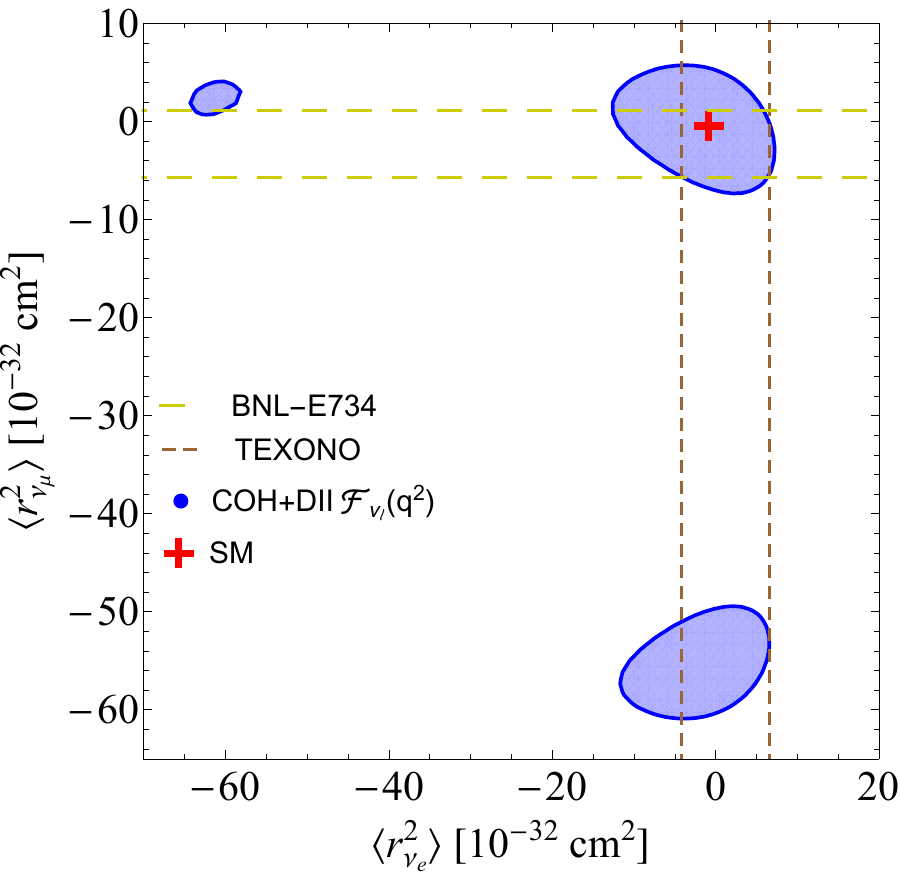}
    \caption{Allowed contours at 90\% CL from the combined analysis of COHERENT and Dresden-II data in the $\langle r^2_{\nu_{e}}\rangle$ vs $\langle r^2_{\nu_{\mu}}\rangle$ plane (allowing only for diagonal NCR contributions). The red cross indicates the SM value of the neutrino charge radii as reported in Eqs.~(\ref{NCReeSM}) and~(\ref{NCRuuSM}). We compare the results with the current best limits on $\langle r^2_{\nu_{\mu}}\rangle$ from BNL-E734~\cite{BNL} (gold) and $\langle r^2_{\nu_{e}}\rangle$ from TEXONO~\cite{TEXONO:2009knm} (brown).}
    \label{fig:COHCsIDIItot}
\end{figure}

By comparing the two figures in Fig.~\ref{fig:COHCsIDII} we observe that the effect of the NCR form factor leads, as expected, to a small shift of the Dresden-II bands, while the closed contours allowed by the COHERENT CsI and Ar data are more significantly affected.
This can be understood considering that COHERENT and Dresden-II data refer to rather different momentum transfer regimes. Namely, the Dresden-II data come from a reactor experiment, which is operated at a much lower energy. 
Moreover, all of the obtained contours are slightly enlarged by the introduction of the form factor. \\
The main impact of accounting for the NCR form factor is that, by combining different measurements, it is possible to significantly reduce the allowed regions in the parameter space.
Indeed, the different momentum transfer regimes relative to the various data, produce a reduced overlap between the allowed regions for non-SM values of $\langle r^2_{\nu_{e}}\rangle$, as shown in Fig.~\ref{fig:COHCsIDIItot}, where the current best limits from BNL-E734~\cite{BNL} on $\langle r^2_{\nu_{\mu}}\rangle$ and TEXONO~\cite{TEXONO:2009knm} on $\langle r^2_{\nu_{e}}\rangle$ are also shown. Interestingly, the SM predicted values for the neutrino charge radii fall within the allowed regions from all the experimental data.\\

\begin{figure}[h]
    \centering
    \includegraphics[width=0.49\textwidth]{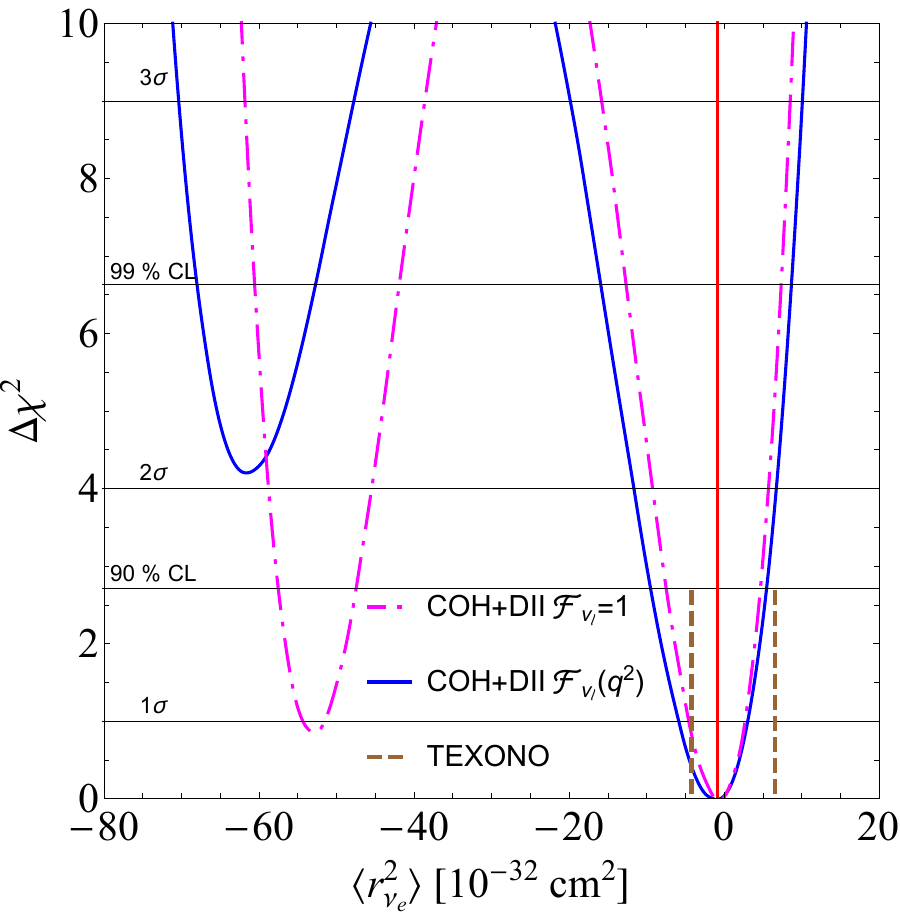}
    \includegraphics[width=0.49\textwidth]{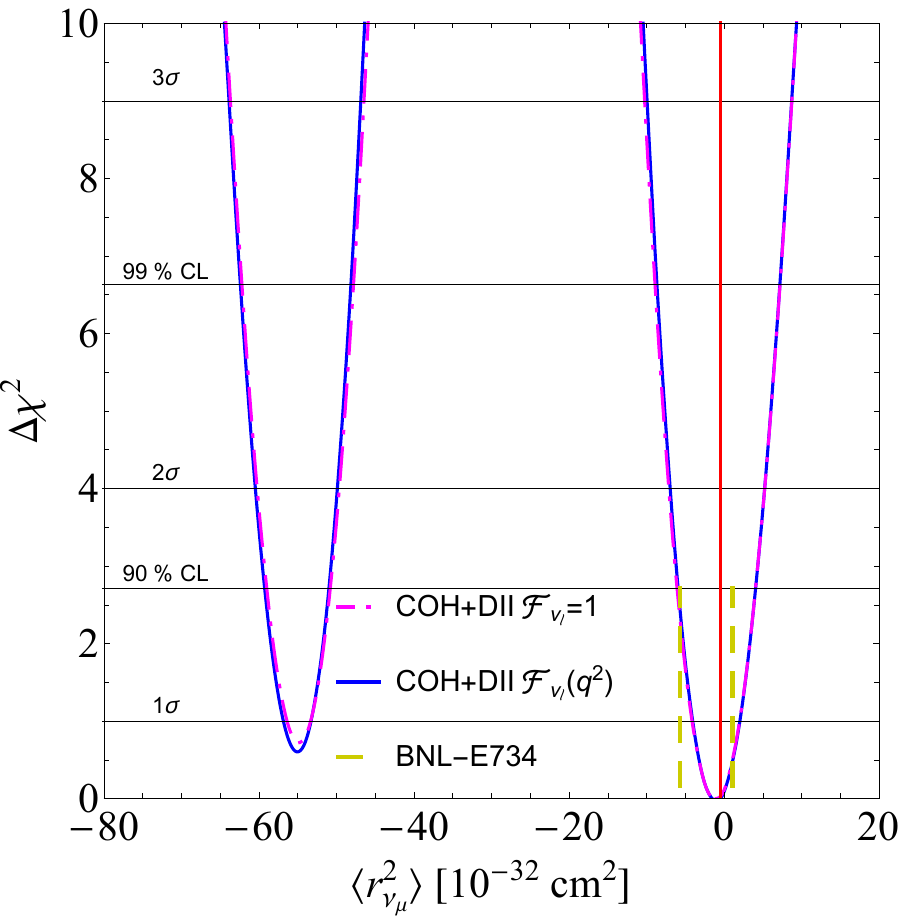}
    \caption{Marginal $\Delta\chi^2$'s
for
$\langle{r}_{\nu_{e}}^2\rangle$ (left) and $\langle{r}_{\nu_{\mu}}^2\rangle$ (right)
obtained from the analysis of the COHERENT (CsI and Ar) data in combination with Dresden-II considering $\mathcal{F}_{\nu_\ell}=1$ (dot-dashed magenta) and $\mathcal{F}_{\nu_\ell}(T_{\rm nr})$ (solid blue).
The red line near the origin indicates the SM predictions.
The short vertical dashed lines show the lower and upper 90\% bounds on
$\langle{r}_{\nu_{e}}^2\rangle$
obtained in the
TEXONO~\cite{TEXONO:2009knm} (left) experiment and on $\langle{r}_{\nu_{\mu}}^2\rangle$ obtained in the BNL-E734~\cite{BNL} (right) experiment.}
    \label{fig:enter-label5}
\end{figure}

Finally, in Fig.~\ref{fig:enter-label5}, we show the marginal $\Delta\chi^2$'s for $\langle r^2_{\nu_{e}}\rangle$ (left) and $\langle r^2_{\nu_{\mu}}\rangle$ (right) obtained from the combined COHERENT and Dresden-II data in the two NCR form factor regimes. 
As expected, the momentum dependence impacts the constraints only for the electron flavor, leaving the results for the muon flavor practically unchanged. In particular, the presence of the NCR form factor decreases the significance of the allowed $\langle r^2_{\nu_{e}}\rangle$ values particularly different from the SM prediction.

The numerical results obtained from our reanalysis are summarized in Tab.~\ref{tab:tab22-chr2-CsI}.
In particular, we obtain a competitive limit on $\langle r^2_{\nu_{e}}\rangle$ at 90\% CL, namely
\begin{equation}
    -9.5<\langle{r_{\nu_{e}}^{2}}\rangle\,[10^{-32}\, \mathrm{cm}^2]<5.5,
    \label{eq:CR_limit}
\end{equation}
and on $\langle r^2_{\nu_{\mu}}\rangle$, 
\begin{equation}
    -59.2<\langle{r_{\nu_{\mu}}^{2}} \rangle\,[10^{-32}\, \mathrm{cm}^2]<-51.0\quad \mathrm{and}\quad-5.9<\langle{r_{\nu_{\mu}}^{2}} \rangle[10^{-32}\, \mathrm{cm}^2]<4.1.
    \label{eq:CR_limit_mu}
\end{equation}
While the bound on $\langle{r_{\nu_{\mu}}^{2}}\rangle$ shown in Fig.~\ref{fig:enter-label5} (right) is practically unchanged with respect to the result for $\mathcal{F}_{\nu_\ell}(T_{\rm nr})=1$ (dot-dashed curve), where no momentum dependence is considered, the constraint set on $\langle r^2_{\nu_{e}}\rangle$ in Fig.~\ref{fig:enter-label5} (left) is largely affected, as the $\Delta\chi^2$'s minimum centred at large negative NCR values is disfavoured at more than $2\sigma$ CL, while it is allowed at about $1\sigma$ CL for $\mathcal{F}_{\nu_\ell}(T_{\rm nr})=1$.
Our results have to be compared to the current best constraints at 90\% CL set by TEXONO~\cite{TEXONO:2009knm}, namely $-4.2<\langle{r_{\nu_{e}}^{2}} \rangle\,[10^{-32}\, \mathrm{cm}^2]<6.6$, and by BNL-E734~\cite{BNL}, $-5.7<\langle{r_{\nu_{\mu}}^{2}} \rangle\,[10^{-32}\, \mathrm{cm}^2]<1.1$.\footnote{Both TEXONO and BNL-E734 results have been corrected by a factor of two due to a different convention, see Ref.~\cite{Cadeddu:2018dux}. Moreover, for the latter, we use the corrected value in Ref.~\cite{Hirsch:2002uv}.} Interestingly, we are able to improve the best upper bound limit for $\langle{r_{\nu_{e}}^{2}} \rangle$ previously set by TEXONO.\\

\section{Conclusions}\label{sec:conclusions}

To conclude, in this work we discuss the need to properly account for the non-null momentum transfer of the experiments in the calculation of the neutrino charge radius radiative correction, focusing in particular on the coherent elastic neutrino-nucleus scattering case. We show that the impact of this correction to the cross-section is small, but non-negligible, especially when considering the contribution of electron neutrinos. Moreover, when aiming to measure the neutrino charge radius, the inclusion of such a momentum dependence results in a significant reduction of the allowed parameter space, already for available data. We thereby stress the importance of using this treatment in particular when analysing future \cenns and neutrino-electron scattering data, for a more accurate interpretation of the data.

\begin{table}
\renewcommand{\arraystretch}{1.2}
\begin{center}
{
\begin{tabular}{ccccc}
&
$1\sigma$
&
$90\%$
&
$2\sigma$
&
$3\sigma$
\\
\hline
\multicolumn{5}{c}{\bf COHERENT (CsI+Ar)}
\\
\multirow{2}{*}{$\langle{r}_{\nu_{e}}^2\rangle$}
&
$( -95.0 , -77.4 )$
&
$( -100.0 , -69.8 )$
&
$( -102.6 , -64.8 )$
&
\multirow{2}{*}{$( -110.4 , 30.7)$}
\\
&
$( 0.09 , 12.8 )$
&
$( -8.6 , 19.1 )$
&
$( -13.9 , 22.2 )$
&
\\
\cline{2-5}
\multirow{2}{*}{$\langle{r}_{\nu_{\mu}}^2\rangle$}
&
\multirow{2}{*}{$( -6.8 , 0.5 )$}
&
$( -57.6 , -48.9 )$
&
$( -59.2 , -47.1 )$

&
$( -63.3 , -42.3 )$
\\
&

&
$( -9.3 , 2.9 )$
&
$( -10.7 , 4.2 )$

&
$( -15.2 , 8.1 )$
\\
\hline
\multicolumn{5}{c}{\bf Dresden-II}
\\
\multirow{2}{*}{$\langle{r}_{\nu_{e}}^2\rangle$}
&
$( -62.5 , -53.7 )$
&
$( -65.7 , -48.5 )$
&
$( -67.2 , -45.0 )$

&
\multirow{2}{*}{$( -71.1 , 9.7 )$}
\\
&
$( -9.0 , 1.8 )$
&
$( -13.8 , 4.5 )$
&
$( -17.2 , 6.0 )$
&

\\
\hline
\multicolumn{5}{c}{\bf COHERENT (CsI+Ar) + Dresden-II}
\\
\multirow{2}{*}{$\langle{r}_{\nu_{e}}^2\rangle$}
&
\multirow{2}{*}{$( -5.8 , 3.1 )$}
&
\multirow{2}{*}{$( -9.5 , 5.5 )$}
&
\multirow{2}{*}{$( -11.6 , 6.8 )$}
&
$( -70.3 , -47.7 )$
\\
&

&

&

&
$( -19.8 , 10.1 )$
\\
\cline{2-5}
\multirow{2}{*}{$\langle{r}_{\nu_{\mu}}^2\rangle$}
&
$( -56.8 , -53.3 )$
&
$( -59.2 , -51.0 )$
&
$( -60.4 , -49.9 )$

&
$( -63.8 , -46.8 )$
\\
&
$( -4.0 , 2.1 )$
&
$( -5.9 , 4.1 )$
&
$( -6.9 ,5.3 )$

&
$( -9.9 , 8.7 )$
\\
\hline
\end{tabular}

}
\end{center}
\caption{ \label{tab:tab22-chr2-CsI}
Bounds on the diagonal neutrino charge radii in units of $10^{-32}~\text{cm}^2$
obtained from our reanalysis of the COHERENT (CsI+Ar) and Dresden-II data alone and their combination at different confidence levels, considering the momentum dependence in the NCR form factor. 
}
\end{table}

\begin{acknowledgements}
The authors are sincerely thankful to J. Erler, M. Gorchtein, H. Spiesberger, R. Ferro-Hernandez and A. Pisano for the fruitful discussions during the completion of this work. 
The work of C. Giunti is supported by the PRIN 2022 research grant  ``Addressing systematic uncertainties in searches for dark matter", Number 2022F2843L, funded by MIUR.

\end{acknowledgements}

\bibliography{ref}

\end{document}